\documentclass[twocolumn,showpacs,prl,superscriptaddress]{revtex4-1}
\usepackage{amsmath,amssymb}
\usepackage{natbib}
\usepackage{comment}
\usepackage{blindtext}
\providecommand{\LyX}{L\kern-.1667em\lower.25em\hbox{Y}\kern-.125emX\@}

\usepackage[T1]{fontenc}
\usepackage[latin1]{inputenc}
\usepackage{graphics}

\usepackage[normalem]{ulem}
\makeatother

\begin{document}

\newcommand{\vK}{\vec{K}}
\newcommand{\vR}{\vec{R}}
\newcommand {\mbf}[1]{{\mathbf{#1}}}
\newcommand {\vecg}[1]{\mbox{\boldmath{$#1$}} }
\newcommand{\be}{\begin{eqnarray}}
\newcommand{\ee}{\end{eqnarray}}

\topmargin -1cm

\title{Interplay between valence and core excitation mechanisms in the breakup of halo nuclei}

\author{A.M.~Moro}
\email{moro@us.es}
\affiliation{Departamento de FAMN, Universidad de Sevilla, Apartado 1065, E-41080 Sevilla, Spain}
\author{J.A.~Lay}
\email{lay@us.es}
\affiliation{Departamento de FAMN, Universidad de Sevilla, Apartado 1065, E-41080 Sevilla, Spain}

\date{\today}

\begin{abstract}
The phenomenon of core excitation in the breakup of a two-body halo nucleus is investigated. We show that this effect plays a significant role in the reaction dynamics and, furthermore, its interference with the valence excitation mechanism has sizable and measurable effects on the breakup angular distributions. These effects have been studied in the resonant breakup of $^{11}$Be on a carbon target,  populating the resonances at 1.78 MeV ($5/2^+$) and 3.41 MeV ($3/2^+$). The calculations  have been performed using a  recently extension of the DWBA method, which  takes into account the effect of  core excitation in both the structure of the halo nucleus and in the reaction mechanism. The calculated angular distributions have been compared with the available data [Fukuda \textit{et al.}, Phys.\ Rev.\ C70,054606]. Although each of these resonances is dominated by one of the two considered mechanisms, the angular patterns of these resonances depend in a very delicate way on the interference between them. This is the first clear evidence of this effect but the phenomenon is likely to occur in other similar reactions. 

\end{abstract}

\maketitle

 {\it Introduction.}--
The study of exotic nuclei has played a key role in nuclear physics over the past 25 years.  
Our current knowledge of their peculiar structural properties comes mainly from measurements of removal cross sections, transfer  and breakup reactions. Breakup reactions have provided useful information on the ground state properties, such as binding energies, spectroscopic factors  and angular momentum (eg.~\cite{Nak09,Nak12}). Moreover, when exclusive measurements are possible, i.e., all outgoing fragments are detected after breakup, these experiments can be used to infer spectroscopic properties of the continuum, such as the location and spin/assignment of resonant states \cite{Aum99,Fuk04,Sat08} and dipole strengths \cite{Nak94,Nak06,Aum99}. 

 In the case of halo nuclei, 
loosely bound exotic nuclei composed of a tightly bound core  surrounded by one or two loosely bound nucleons, these processes have been  conveniently modeled using a three-body model, comprising the two-body weakly bound projectile and the target. 
 This  has motivated  the development and revival of few-body theories, such as the  Continuum--Discretized Coupled--Channels (CDCC) method \cite{CDCC}, the adiabatic ({\it frozen-halo}) approximation \cite{JS70,Ron97,Cres99}, a variety of semiclassical approaches \cite{Typ94,Esb96,Kid94,Cap04} and, more recently, 
the Faddeev equations \cite{faddeev60,Alt,deltuva07,Cravo10}.

 In their standard formulations, these methods assume a single-particle description of the valence particle relative to the core. 
 Possible excitations of the core are neglected or, at most, taken into account effectively through the core-target optical potential.  Although this approach has been used with relative success in the analysis of many reactions, it has been recently shown \cite{Crespo11b,Mor12} that this simplified picture is not always accurate, due to the effects of  core excitation. This affects in two ways. First, the presence of core admixtures in the states of the projectile means that these states cannot be simply treated as single-particle states calculated in some mean field potential. Second, the interaction of the core with the target may give rise to transitions between these core states, leading also to the breakup of the projectile. Although these two effects have been commonly ignored in the analysis of reactions with exotic beams, some
 progress has been made in  recent years toward their incorporation in existing reaction formalisms. 
 For example,  Summers {\it et al.} \cite{Summers06} have proposed an extended version of the CDCC method which treats the 
structure of the few-body projectile within the particle--rotor model.  More recently, a simple extension of the standard DWBA amplitude
which takes into account these effects approximately has been proposed \cite{Crespo11b,Mor12}. Using this method, it was found that dynamic core excitation are indeed very important to describe the breakup of $^{11}$Be on protons at $\sim$70~MeV/nucleon.  Some effects have been also found in the elastic scattering of  $^8$B on a carbon target \cite{Hor10}, using an extension of the adiabatic model of Ref.~\cite{Ron97}.

Although the calculations presented in \cite{Crespo11b,Mor12} provide a clear evidence of the importance of core excitation in nuclear breakup, the restricted energy and angular resolution of the analysed data prevented  a detailed assessment of the relative importance and the interplay between the valence and core excitation mechanisms. 
In this Letter, we present new  results showing that the presence of core admixtures in the halo nucleus and the subsequent dynamic core transitions give rise to very distinctive effects on the shape and magnitude of the  breakup cross sections angular distributions. Moreover, it is found that these effects depend very critically on the amount of core excitation of the halo nucleus. This sensitivity opens new possibilities for extracting spectroscopic information of halo and other weakly bound nuclei, by comparing the measured angular distributions with a reliable reaction model.  To illustrate these effects, we present here calculations for the 
breakup of the one-neutron halo nucleus ${\rm ^{11}Be}$ on a carbon target at  a bombarding energy of 70 MeV/nucleon \cite{Fuk04}, in particular to the excitation of the resonances at $E_x=1.78$~MeV and 3.41 MeV. 
The calculations are performed using the extended version of the DWBA method of Refs.~\cite{Crespo11b,Mor12}.   


 {\it Reaction model}.-- 
We now describe briefly the core-excitation model used here. We outline here the main 
formulas, and refer the reader to Refs.~\cite{Crespo11b,Mor12} for further details. 
We consider the inelastic excitation of a projectile nucleus, initially in its ground state,  $\Psi^i_{J M }$, to a  state $\Psi^f_{J' M'}$ (bound or unbound). Within a two-body (core+valence) description of the projectile, these wavefunctions are expanded as
\begin{equation}
\Psi_{J M }(\vec{r},\vec{\xi}) 
 =  \sum_{\alpha}  
\left[   \varphi_\alpha(\vec{r}) \otimes \Phi_{I}(\vec{\xi}) \right]_{JM} ,
\label{wfdef}
\end{equation}
where the functions $\varphi_\alpha(\vec{r})$ describe the relative motion between the valence particle and the core
and  $\Phi_{I M_c}(\vec{\xi})$  are the core eigenstates  with angular momentum $I$ and projection $M_c$. The index  $\alpha$  denotes the set of quantum numbers $\{\ell,j,I \}$, with   
 $\ell$, $s$ and $j$ being  the  orbital angular momentum, the intrinsic spin of the valence 
particle and their sum ($\vec{j}=\vec{\ell} + \vec{s}$), respectively. 
 The functions $\varphi_\alpha(\vec{r})$ and $\Phi_{I M_c}(\vec{\xi})$  depend on the assumed structure model, to  be specified later. 

Consistently with the assumed two-body description of the projectile, the transition
potential for this breakup process is the sum of the valence-target and core-target interactions, i.e., $V_T= V_{vt}(\vec{R}_{vt}) + V_{ct}(\vec{R}_{ct},\vec{\xi})$. While the valence-target interaction is taken to be central and to depend exclusively on the valence-target separation, the core-target interaction is assumed to depend  on the core internal degrees of freedom, and can therefore induce transitions between different core states. By expanding this interaction in multipoles ($\lambda$) and separating the central ($\lambda$=0) part, the DWBA amplitude, describing the excitation of the halo nucleus during the collision, splits into two terms:
\be
T^{J M,J' M'}(\vec{K}',\vec{K})=T^{J M,J' M'}_{val} + T^{J M,J' M'}_\mathrm{corex},
\label{dwba}
\ee
 with 
\begin{align}
T^{J M,J' M'}_{val}(\vec{K}',\vec{K}) & =
\langle \chi^{(-)}_{\vK'}(\vec{R}) \Psi^f_{J' M'}(\vec{r},\vec{\xi}) |
 V_{vt}(R_{vt})   \nonumber \\
&  + V^{(0)}_{ct}( R_{ct})| 
\chi^{(+)}_{\vK}(\vec{R}) \Psi^i_{J M}(\vec{r},\vec{\xi})  \rangle \,
 \label{dwba_sp}
\end{align}
where $\vec{K}$ ($\vec{K}'$) is the initial (final) linear momentum and $\chi^{(+)}_i(\vec{R})$  ($\chi^{(+)}_f(\vec{R})$)  the initial (final) distorted wave describing the projectile-target relative motion.
The transition amplitude given by Eq.~(\ref{dwba_sp}) ({\it valence amplitude} hereafter) 
describes excitations between different valence configurations, but without altering the state of the core. This term is evaluated following the 
standard techniques used in coupled-channels codes. 
The  second term in (\ref{dwba}), denoted  {\it core excitation amplitude},  contains the non-central 
part ($\lambda$>0) of the core-target interaction  and can therefore produce core transitions.
Consequently, this term accounts for the dynamic excitation of the core during the collision.  In \cite{Crespo11b,Mor12}, it was shown that this  
term acquires a very simple form when evaluated in the no-recoil approximation  ($\vec{R}_{ct} \approx \vec{R}$), 
\begin{align}
 T_{\rm corex}^{J M,J' M'} &=  \sum_{\lambda>0, \mu} \langle J' M'|J M \lambda \mu \rangle \nonumber \\
&\times     \sum_{\alpha, \alpha'}  \langle R^{J'}_{\alpha'} | R^{J}_\alpha \rangle 
 G^{\lambda}_{\alpha J,\alpha' J'} {\widetilde T}_{\rm ct}^{(\lambda \mu)}(I\rightarrow I')
\label{dwba_core}
\end{align}
where $G^{(\lambda)}_{\alpha J,\alpha' J'} $ is a geometric factor \cite{Crespo11b,Mor12},  $R^{J}_{\alpha}$ are the radial parts of the 
$\varphi_\alpha(\vec{r})$ functions 
and  $\widetilde{T}^{(\lambda \mu)}_{ct}$ is related to the  core-target two-body transition amplitude for 
a core transition $I M_c \rightarrow I' M'_c$ of multipolarity $\lambda$ as 
$T^{I M_c, I' M'_c}_{ct}=  \langle I M_c \lambda \mu | I'M'_c \rangle  \widetilde{T}^{(\lambda \mu)}_{ct}$.

{\it Results.--} The model has been  applied to the resonant breakup of $^{11}$Be on a $^{12}$C target at 70 MeV/nucleon, measured at the RIKEN facility 
by Fukuda {\it el al.} \cite{Fuk04}. The relative energy spectrum
of the $^{10}$Be+$n$ center of mass 
shows peaks at $E_x$=1.78 MeV and $E_x$=3.41 MeV. Their angular distributions were compared with DWBA calculations, based on the vibrational collective model, suggesting a $\lambda=2$ transition for both states. These states were identified with the  $5/2^+_1$ and $3/2^+_1$ resonances predicted by shell-model calculations.

In the present work, we reanalyze these data using the aforementioned core-excitation DWBA model. The structure of the 
$^{11}$Be nucleus is described within the particle-rotor
model (PRM)  of Bohr and Mottelson, with the parameters given by the model Be11b of
Ref.~\cite{Nun96b}. This model assumes  a permanent quadrupole deformation for the  $^{10}{\rm Be}$ core
with $\beta_2$=0.67.

 The $^{11}$Be wave functions are obtained by diagonalizing the $^{11}$Be Hamiltonian 
in a particle+core basis of the form $|\Phi_{I}(\vec{\xi})\otimes \varphi^\mathrm{THO}_{n (\ell s)j}(\vec{r})\rangle_{JM}$,  
where $\varphi^\mathrm{THO}_{n (\ell s)j}(\vec{r})$ (with ~n=1,\ldots,N) are a truncated set of Transformed Harmonic Oscillator (THO) functions \cite{Lay12}, which are used as a basis for the valence-target relative motion. This basis is obtained by applying an analytic  local scale transformation (LST) to the conventional harmonic oscillator basis. The modelspace is restricted to $I^\pi=0^+,2^+$ and  $\ell \leq 2$. For
the ground state, a basis of N=15 oscillator functions was used. 
Resonant states are identified with stabilized energies with respect to the basis size (N). The parameters of the LST are the same as those used in \cite{Lay12}. 

The components involved in our calculations allowed by the present model space,  along with their respective weights (spectroscopic factors) 
are listed in Table \ref{Tab:sf}.  Also listed in this table are the shell-model spectroscopic factors obtained with the code {\sc oxbash}, using the WBT interaction of Warburton and Brown \cite{wbt}. Both models  predict very similar spectroscopic factors for the ground state and the $5/2^+$ resonance, and only some small differences are found in the $3/2^+$ state. 
The ground state corresponds  predominantly to a
$ | ^{10}{\rm Be}(0^+) \otimes  s_{1/2} \rangle $  
configuration,  with some 
admixture of the  $ | ^{10}{\rm Be}(2^{+})  \otimes d_{5/2} \rangle $ configuration. The $5/2^+$ state is mainly based on the $^{10}$Be ground state. 
 On the other hand, 
the $3/2^+$  resonance is mainly built on top of the excited core. According to this result, it is expected that the population of the $5/2^+$ state is mainly due to the valence excitation mechanism, whereas the excitation of the $3/2^+$ state will be mostly due to a core-excitation mechanism.

 To illustrate the sensitivity of the calculation with the structure model, we have considered two additional models assuming pure single-particle 
configurations for the $^{11}$Be g.s.\ and the  
$5/2^+$ and $3/2^+$ resonances. For the $^{11}$Be(g.s.) we consider a pure $|0^+ \otimes 2s_{1/2}\rangle$ configuration. For the $5/2^+$ resonance we consider two single-particle models: i)  $|0^+ \otimes 1d_{5/2}\rangle$  (denoted SP1) and ii)   $|2^+ \otimes 2s_{1/2}\rangle$  (SP2). In the former,  the resonance is populated by means of a valence excitation mechanism, whereas in the second model the excitation is due to a pure core excitation effect. Similarly, for the 
$3/2^+$ we consider also two extreme models: i) $|0^{+} \otimes 1d_{3/2}\rangle$  (SP1) and ii)   $|2^{+} \otimes 2s_{1/2}\rangle$  (SP2). The required radial wavefunctions are taken from the PRM calculation, conveniently normalized to one.

\begin{table*}[!ht]
\caption{\label{Tab:sf} Spectroscopic factors for the ground state and resonant wavefunctions of $^{11}$Be, according to the particle-rotor  model (PRM) and the shell-model calculations (WBT) presented in this work.}
\begin{center}
\begin{tabular}{ccccccc}
\hline
State & Model & $| 0^+ \otimes (\ell s)j \rangle $ &    $ | 2^+ \otimes s_{1/2} \rangle $  &   $ | 2^+ \otimes d_{3/2} \rangle $ & $ | 2^+ \otimes d_{5/2} \rangle $  &  \\
\hline  
\hline
 $1/2^+$ (g.s.)         &  PRM    & 0.857    &    --   &     0.021              &    0.121    \\
               &  WBT    & 0.762    &      -- &               0.002          & 0.184  \\  
\hline
 $5/2^+$ ($E_x$=1.78 MeV)                          &   PRM     &  0.702   &     0.177 & 0.009 & 0.112  \\ 
                        &      WBT     &  0.682  &   0.177          &  0.009      &           0.095       \\ 
\hline
 $3/2^+$ ($E_x$=3.41 MeV)                           &      PRM        & 0.165 &  0.737  & 0.017  & 0.081   \\  
                          &     WBT     &  0.068   &   0.534   &  0.008   &      0.167  \\
\hline
\end{tabular}
\end{center}
\end{table*}

The $n$+$^{12}$C potential was taken from Ref.~\cite{Bec69}. The central and transition components of the $^{10}$Be+$^{12}$C potential 
were generated by a double folding procedure, convoluting an effective nucleon-nucleon (NN) interaction with the $^{10}$Be and $^{12}$C matter densities. The 
latter were taken, respectively, from the Antisymetrized Molecular Dynamics (AMD) calculation of Ref.~\cite{Kan02} and from the parametrization of Ref.~\cite{Ela85}. For the effective NN interaction we adopt the
spin-isospin independent part of the M3Y interaction \cite{M3Y} based on the Reid soft-core NN potential. For the imaginary part of the $^{10}$Be+$^{12}$C potential we assume the 
same geometry as for the real part. A renormalization factor was included to reproduce the elastic scattering data of  $^{10}$Be+$^{12}$C  at 59.4 MeV/nucleon
from Ref.~\cite{Cor96,*Cor97}. Further details of these calculations will be provided elsewhere. 


\begin{figure}
\resizebox*{0.85\columnwidth}{!}
{\includegraphics{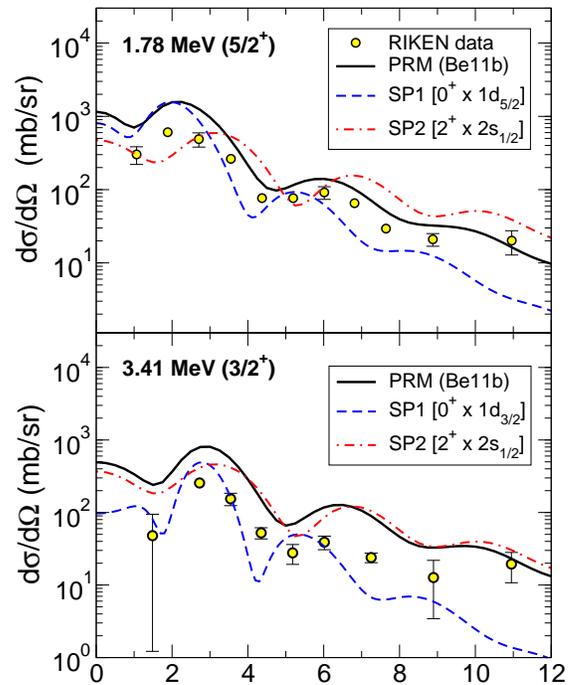}}
\caption{\label{dsdw_comp}
(Color online) Angular distribution for the $E_x$=1.78 MeV and 3.41 MeV states in $^{11}$Be. The circles are the data from Ref.~\cite{Fuk04}. The curves correspond to the extended DWBA calculations, including core excitation effects, using different structure models for the  $^{11}$Be nucleus. For the single-particle models (SP1 and SP2) the resonance configuration is indicated in the labels.}
\end{figure}

In Fig.~\ref{dsdw_comp} we compare the
calculated  angular distributions with the  experimental data of Ref.~\cite{Fuk04}. The upper and bottom panels correspond to the $5/2^+$ ($E_x$=1.78 MeV) and $3/2^+$ ($E_x$=3.41 MeV) resonances. 
It is readily seen that the pure single-particle models SP1 and SP2 do not reproduce the shape of the resonances. In the model SP1 (pure valence excitation) 
the maxima and minima are shifted to smaller angles with respect to the data and the angular distribution decay too fast. On the other hand, in the model SP2
(pure core excitation mechanism) the maxima and minima are shifted to larger angles. Finally, the full PRM model, which includes both valence and core excitation mechanisms and their interference, the position of the maxima and minima is very well reproduced. It is also seen that the absolute magnitude of the data is overestimated.  Except for this discrepancy in the normalization, it is clear that the shape is appreciably improved with respect to the pure single-particle  description and that the interference between the  valence and core excitation mechanisms is crucial to account for the correct shape of the oscillations. 
\begin{figure}
\resizebox*{0.9\columnwidth}{!}
{\includegraphics{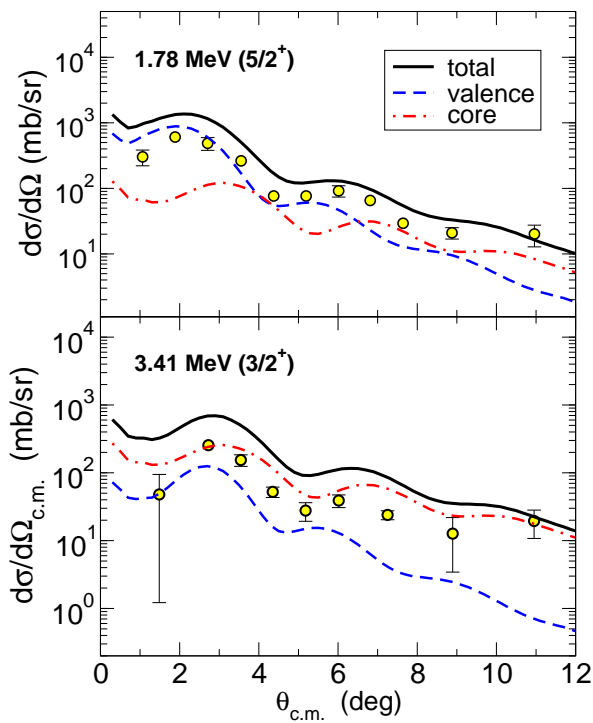}}
\caption{\label{dsdw_vc}
(Color online) Valence (dashed) and core (dot-dashed)  contributions to the breakup of the 1.78 and 3.41 MeV resonances populated in the $^{11}$Be+$^{12}$C reaction at 70 MeV/nucleon, using a particle-core description of the $^{11}$Be nucleus. The solid is the coherent sum of both contributions. }
\end{figure}

It is enlightening to consider separately the contribution of the valence and core excitation amplitudes, 
Eqs.~(\ref{dwba_sp}) and (\ref{dwba_core}). These are depicted in Fig.~\ref{dsdw_vc} for the PRM model. 
 In this plot,  the calculations have been convoluted with the experimental angular  resolution \cite{Fuk04} for a more meaningful comparison with the data.
As anticipated,  the $5/2^+$ resonance is mainly populated by the valence excitation mechanism, due to its dominant $^{10}{\rm Be(0^+)}$ configuration, whereas for the $3/2^+$ state the dynamic core excitation mechanism is the dominant one. It is also seen that both contributions are out of phase, and  the interference between them is very important. In fact, none of them separately is able to reproduce by itself the position of the maxima and minima of the data, whereas their coherent sum (solid line) reproduces very well this pattern. This result illustrates very nicely the delicate interplay between the valence and core excitation mechanisms in the breakup of a deformed halo nucleus, like  $^{11}{\rm Be}$. Note that the weak contribution of the valence mechanism in the $3/2^+$ case is a consequence of the small spectroscopic factor 
associated to the $|0^+ \otimes d_{3/2}\rangle$ configuration (see Table \ref{Tab:sf}). This fact explains also that this resonance is very 
weakly populated in transfer reactions, such as $^{10}{\rm Be}(d,p)^{11}{\rm Be}$ \cite{Zwi79,*Del03}, making it difficult the  extraction of spectroscopic information from these experiments. In these cases, the approach presented in this work, based on  the analysis of breakup reactions, provides a powerful alternative to access this information.

{\em Conclusions.}--
In conclusion, we have studied the interplay between the valence and core excitation mechanisms in the breakup of halo nuclei using and recently proposed extension of the DWBA method. We have shown that the presence of core admixtures in the 
initial and final states has a sizable impact in the interference pattern of the breakup cross section and hence a high sensitivity on the underlying structure model of the halo nucleus. This effect has been evidenced for the first time in the scattering of $^{11}$Be on $^{12}$C  at 70 MeV/nucleon, where we have shown that the inclusion of these core excitation effects improves significantly the agreement with the data \cite{Fuk04} and provides very valuable spectroscopic information, which would be very difficult to extract from other methods.  Finally, we emphasize that, although the calculations have been presented for the 
$^{11}$Be nucleus, we do expect  these effects to be important in other relevant  cases, such as in the breakup of the odd carbon isotopes $^{15,17,19}{\rm C}$. 




\begin{acknowledgments}
  We are grateful to  Dr.~Y. Kanada En'yo for providing us the 
$^{10}$Be microscopic densities and to T.\ Nakamura for his help regarding the $^{11}$Be+$^{12}$C data and the convolution with the experimental resolution. 
This work has been partially supported by the   Spanish Ministerio de  Ciencia e Innovaci\'on under project FPA2009-07653, and 
by the Spanish Consolider-Ingenio 2010 Programme CPAN (CSD2007-00042).  J.A.L.\ acknowledges a research grant by the Ministerio de Ciencia e Innovaci\'on.
\end{acknowledgments}

\bibliographystyle{apsrev4-1}
\bibliography{be11c}

\end{document}